\documentclass[aps,pra,reprint,superscriptaddress,floatfix]{revtex4-1}

\usepackage{amsmath,amssymb,amsthm,cases,graphicx,bm}
\usepackage[utf8]{inputenc}
\usepackage{dcolumn}
\usepackage[colorlinks=true,citecolor=blue,urlcolor=blue]{hyperref}
\usepackage[dvipsnames]{xcolor}
\usepackage{subfigure}
\usepackage{soul}
\newcommand{\eqnref}

\begin{document}

\title{
Effect of interactions in the interference pattern of Bose-Einstein condensates}

\author{A.~Burchianti}
\affiliation{Istituto Nazionale di Ottica, CNR-INO, 50019 Sesto Fiorentino, Italy}
\affiliation{\mbox{LENS and Dipartimento di Fisica e Astronomia, Universit\`{a} di Firenze, 50019 Sesto Fiorentino, Italy}}

\author{C.~D'Errico}
\affiliation{Istituto per la Protezione Sostenibile delle Piante, CNR-IPSP, 10135 Torino, Italy}

\author{L. Marconi}
\affiliation{Istituto Nazionale di Ottica, CNR-INO, 50019 Sesto Fiorentino, Italy}

\author{F. Minardi}
\affiliation{Istituto Nazionale di Ottica, CNR-INO, 50019 Sesto Fiorentino, Italy}
\affiliation{\mbox{LENS and Dipartimento di Fisica e Astronomia, Universit\`{a} di Firenze, 50019 Sesto Fiorentino, Italy}}
\affiliation{Dipartimento di Fisica e Astronomia, Universit\`{a} di Bologna, 40127 Bologna, Italy}

\author{C.~Fort}
\affiliation{Istituto Nazionale di Ottica, CNR-INO, 50019 Sesto Fiorentino, Italy}
\affiliation{\mbox{LENS and Dipartimento di Fisica e Astronomia, Universit\`{a} di Firenze, 50019 Sesto Fiorentino, Italy}}

\author{M. Modugno}
\affiliation{\mbox{
Department of Physics, University of the Basque Country UPV/EHU, 48080 Bilbao, Spain}}
\affiliation{IKERBASQUE, Basque Foundation for Science, 48013 Bilbao, Spain}

\begin{abstract}
Understanding the effect of interactions in the phase evolution of expanding atomic Bose Einstein condensates is fundamental to describing the basic phenomenon of matter wave interference.
Many theoretical and experimental works tackled this problem, always with the implicit assumption that the mutual interaction between two expanding condensates \textit{rigidly} modifies the phase evolution through an effective force. In this paper, we present a combined experimental and theoretical investigation of the interference profile of expanding $^{87}$Rb condensates, with a specific focus on the effect of interactions.
We come to the different conclusion that the mutual interaction produces \textit{local} modifications of the condensate phase only in the region where the wavepackets overlap.
\end{abstract}

\maketitle

The first experimental evidence of the interference between two atomic Bose Einstein condensates back in 1997 \cite{andrews1997} was welcomed as a breakthrough demonstration of macroscopic phase coherence.
Since then, interference of two or multiple condensates has been the focus of intense research, both to study the coherence properties \cite{simsarian2000}, and to detect the presence of phase defects, such as vortices or solitons \cite{Aidelsburger2017,Serafini2017} and even to reveal spin-orbit coupling \cite{Mardonov2014}. In addition, condensates were soon recognized as ideal sources for matter-wave interferometry \cite{Torii2000}, and used to measure gravity \cite{Debs2011,Hardman2016,Abend2016}, rotations \cite{Marti2015,Navez2016,Pandey2019}, and fundamental physical constants \cite{Gupta2002,Jamison2014}.
In this context interatomic interactions play an important role: on one hand, they might induce detrimental dephasing \cite{Olshanii2005,Horikoshi2006,Jannin2015}; on the other hand, they are a resource for entanglement-enhanced sensitivity, e.g., via the use of squeezed states  \cite{Pezze2005,Lee2006,Huang2008,Gross2010,Berrada2013}.

Despite repeated scrutiny \cite{rohrl1997,liu2000,benton2011,jamison2011}, in the presence of interactions even the basic phenomenon of two interfering condensates still shows interesting and unraveled features \cite{Watanabe2012,Yang2019,Tang2019}. At the mean field level, the self-interaction of an individual condensate drives its phase evolution \cite{castin1996}, while the mutual interactions between two condensates are usually taken into account via the modification of the condensates center-of-mass motion \cite{simsarian2000,benton2011,guan2020}. In doing so, the  implicit assumption that the condensate phase is ``rigid'' \cite{Anderson1994,Sols1998} is made, i.e., not locally deformable but only globally variable.

Here we show, through an experiment backed by a theoretical analysis and numerical simulations, that the assumption of phase rigidity must be abandoned, at least in certain regimes. Indeed, we find
that the mutual interactions between two interfering condensates lead to local modifications of their phase that are essential for the precise description of the condensates interference.
These results are obtained by performing a quantitative analysis of the interference pattern of interacting condensates in free fall and comparing the experimental findings with numerical simulations based on the Gross-Pitaevskii (GP) equation and with semianalytical models that assume phase rigidity. We show that while  semi-analytic models fail to quantitatively describe the observed fringe spacing, these are instead well reproduced by the GP equation. Also, we show that the phase of the macroscopic wave function cannot be described only through the dynamical variables of the center-of-mass motion, i.e., velocity and position.

The paper is organized as follows. Section~\ref{sec:experiment} presents the experimental results obtained applying a sequence of Bragg pulses on an expanding $^{87}$Rb condensate. The integrated density distribution of the interferogram is analyzed performing a Fourier transform (FT). The main
wave vector obtained from the Fourier analysis is compared with (i) the analytical expression for two expanding condensates neglecting their mutual interaction, (ii) the prediction including mutual interaction through an effective force \cite{guan2020}, and (iii) with numerical GP simulations. The measurements clearly show that the mutual interaction between the two expanding condensates modifies their phase evolution.
In Sec.~\ref{sec:model} we  present a theoretical analysis of the simplest scenario of two interfering condensates \cite{andrews1997,naraschewski1996}, revealing that the phase modification due to mutual interactions is not fully captured by an effective force arising from the repulsion between the two wave packets, as considered for example in \cite{simsarian2000,benton2011,guan2020}.
In particular, we evidence a local modification of the phase in the region where the two condensates superimpose.
Finally, in Sec.~\ref{sec:SO} we summarize our results and discuss the outlooks.

\section{Experiment}
\label{sec:experiment}
\begin{figure}[b!]
\begin{center}
\includegraphics[width= 1\columnwidth]{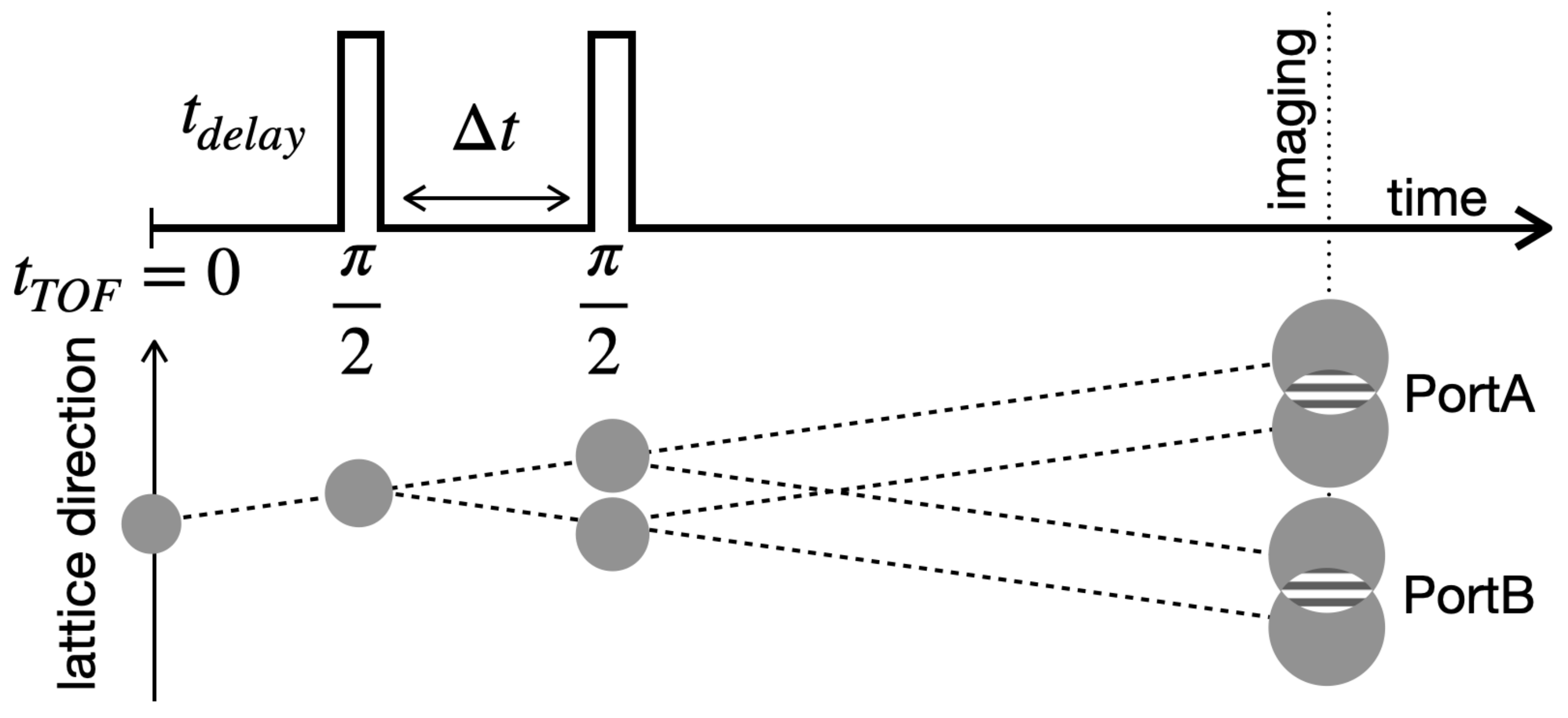}
\end{center}
\caption{Schematic of the interferometric procedure.
}\label{schema_interferometro}
\end{figure}

Expanding interacting condensates are produced with an interferometric sequence of two resonant $\pi/2$ Bragg pulses of a lattice potential, separated by a time interval $\Delta t$. Each $\pi/2$ Bragg pulse is a matter-wave \textit{beam splitter}, coupling the two $\pm\hbar k_L$ momentum states along the lattice direction, where $k_L= \pi/ d$ and $d$ is the optical lattice constant \footnote{A three pulse sequence ($\pi/2-\pi-\pi/2$) is often used \cite{simsarian2000}, where the $\pi$ pulse acts as a mirror in analogy with the Mach-Zehnder optical counterpart.}.
After the second pulse, at both output ports of the interferometer we have two expanding and interfering condensates, which have started their expansion separated by the distance accumulated in the time interval between the two pulses, approximately $2(\hbar k_L/m)\Delta t$, where $m$ is the atomic mass (see Fig.~\ref{schema_interferometro}).

\begin{figure}[t!]
\begin{center}
\includegraphics[width= 0.7\columnwidth]{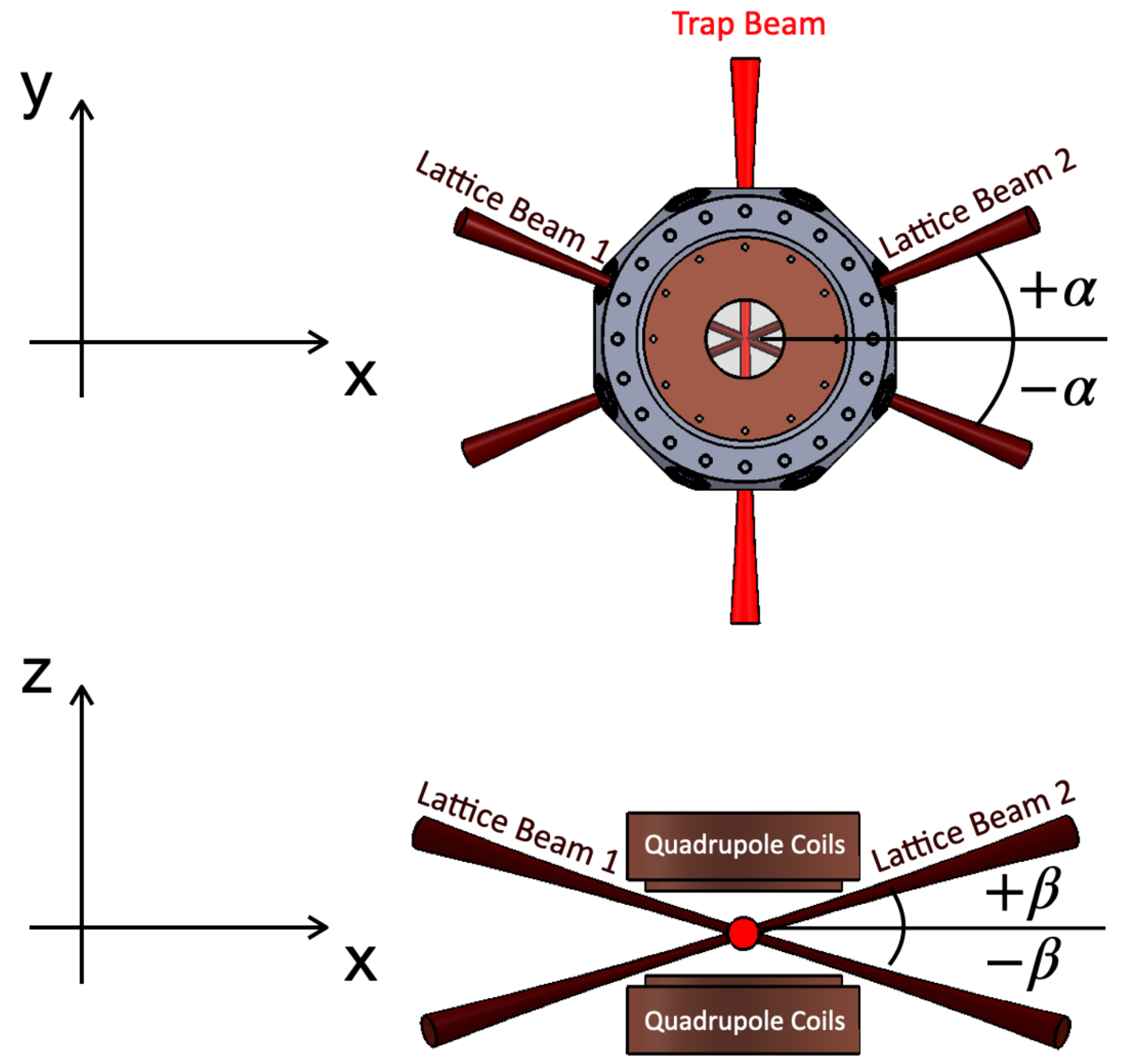}
\end{center}
\caption{Schematic of the experimental apparatus, showing the beam of the optical dipole trap (red), the quadrupole coils (light brown), and the beams of the optical lattice used for the Bragg pulses (brown). Both the trap and the lattice beams are at a wavelength $\lambda_{laser} =1064$~nm.}
\label{schema_apparato}
\end{figure}

In particular, we produce a single $^{87}$Rb condensate with $N=2 \cdot 10^5 - 5 \cdot 10^5$ atoms in the $|F=2,m_F=2\rangle$ state in a hybrid trap \cite{Spielman2009} using a quadrupole magnetic field and an optical dipole potential generated by a single focused laser beam, as detailed in Ref. \cite{Burchianti2018} and schematically shown in Fig.~\ref{schema_apparato}. Typical frequencies of the confining potential are $(\nu_x,\nu_y,\nu_z)\simeq(50,15,60)$~Hz.
We then excite a dipole motion along the $x$ direction. When the condensate reaches the velocity $v=\hbar k_L/m$, we switch off the trapping potential and, after a delay time $t_{delay}$, we perform the Bragg pulse sequence.

The optical lattice potential for the Bragg pulses is produced by two laser beams at $\lambda_{laser}=1064$~nm propagating (as schematically shown in Fig.~\ref{schema_apparato}) at angles  $-\alpha=-22.5^\circ$ (Lattice Beam 1) and $\pi+\alpha$ (Lattice Beam 2) with respect to the $x$ axis in the horizontal $xy$ plane, and inclined at angles of $-\beta=-16^\circ$and $\pi+\beta$, with respect to the same plane. This configuration produces an optical lattice along the $x$ axis with a spacing $d=\lambda_{laser}/(2\cos \alpha \cdot \cos \beta)=599$~nm and $k_L=\pi/d$.
 We typically have a lattice potential height $V_{OL}\simeq 5 E_r$, where $E_r=\hbar^2 k_L^2/(2 m)$ and $m$ are the recoil energy and the mass for $^{87}$Rb atoms.
The Bragg pulse, lasting $t_{pulse}\approx 65$~$\mu$s, is a $\pi/2$ pulse, i.e., a beam-splitter producing two equally populated condensates with opposite momenta $\pm\hbar k_L$.
After the two Bragg pulses, we measure the atomic density distribution waiting an additional time interval adjusted to keep constant the total time of flight $t_{TOF}=33$~ms. At each output port of the interferometer, labeled with A and B as indicated in the schematic of Fig.~\ref{schema_interferometro}, we observe the interference of two condensates that are prepared by the interferometric sequence in a configuration where they are spatially separated and have (approximately) the same velocity.

We performed two different sets of measurements:
in the first set we vary the condensate separation $\Delta x$ by varying the time $\Delta t$ between the two Bragg pulses and keep constant the delay before the first Bragg pulse, $t_{delay}$; in the second set, vice versa, $\Delta t$ is fixed and $t_{delay}$ is varied.

\begin{figure}[th!]
\begin{center}
\includegraphics[width= 1\columnwidth]{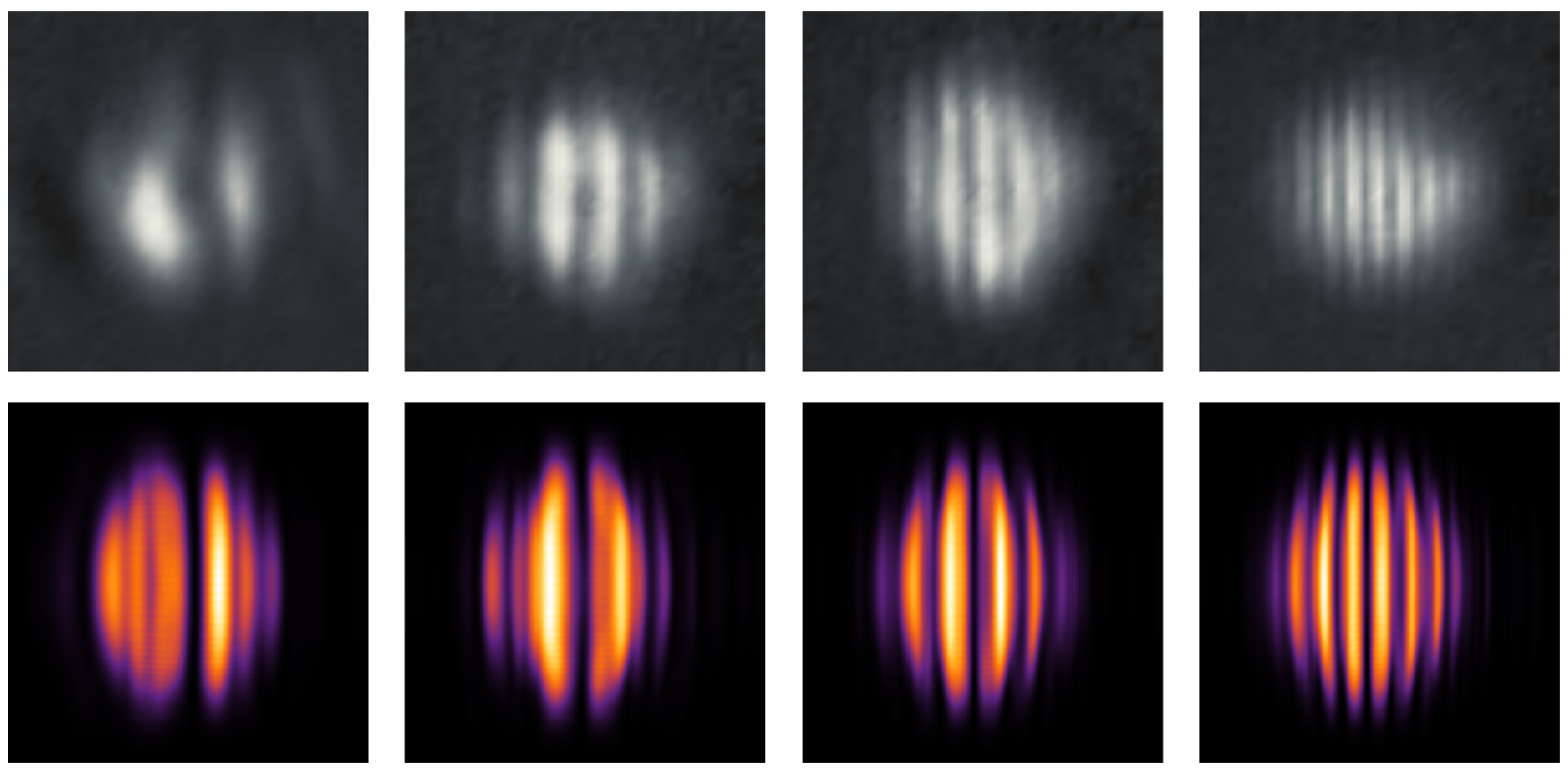}
\end{center}
\caption{Column density in the $xz$ plane for $t_{delay}=1$~ms and different values of $\Delta t$ (0.6~ms, 1.0~ms, 1.4~ms, and 1.8~ms, from left to right). The mean atom number is $N=2\cdot 10^5$ and the harmonic trapping frequencies
(of the initial trap)
are $(\omega_x, \omega_y, \omega_z)\simeq 2\pi~\times~(50,15,60)$~Hz. First row: density distribution measured at the port B of the interferometer.
Second row: GP simulation for the same experimental parameters. The box size is $192~\mu$m.}
\label{free_space_Dt}
\end{figure}
Let us start by the former set.
In Fig.~\ref{free_space_Dt} upper row, we show the measured atomic density distribution integrated along the $y$ direction measured at port B of the interferometer with $t_{delay}=1$~ms varying the time between the two Bragg pulses $\Delta t$ from 0.6~ms to 1.8~ms.
Note that the contrast of the interferometer at port B is expected to be $C=1$ independent from fluctuations in the efficiency of the Bragg coupling, as long as the two pulses are equal  \footnote{With $\epsilon$ being the efficiency of the Bragg pulse, the populations of the two interfering condensates at port A are $N_{A1}=N \epsilon^2$ and $N_{A2}=N(1-\epsilon)^2$, while at port B are $N_{B1}=N_{B2}= N\epsilon(1-\epsilon)$, respectively. As a consequence the contrast of the port A interferogram is $C_A=2\epsilon(1-\epsilon)/[\epsilon^2+(1-\epsilon)^2]$ and varies with $\epsilon$, while $C_B=1$ independently from $\epsilon$.}.
We typically measure a smaller contrast, $C\approx0.4$, at both ports, which we attribute to an angle  of $\approx 5^\circ$ between the detection view, defined by the direction of probe laser beam, and the $yz$ plane of the fringes. This angle affects the fringes' spacing negligibly ($0.3\%$).
The lower row of Fig.~\ref{free_space_Dt} shows the corresponding density profiles obtained from the GP simulation.

\begin{figure}[th!]
\begin{center}
\includegraphics[width= 0.9\columnwidth]{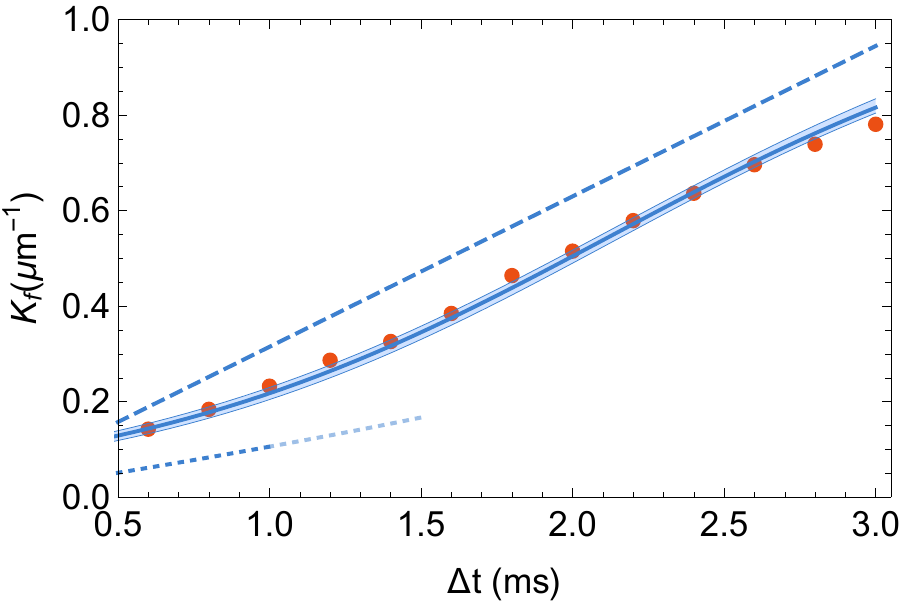}
\end{center}
\caption{Fringe wave vector $K_f$ as a function of $\Delta t$:
the experimental data (red points) are compared with the results of the GP simulations (continuous blue line with a band representing a 50\% uncertainty on $N$) and the theoretical prediction of Eq.~(\ref{lambda_teo}) for the noninteracting case (dashed blue line). The dotted blue line is calculated correcting the center-of-mass motion of the two condensates using the effective force in Eq. (\ref{eq:Fx}) (see text). Here  $t_{delay}=1$~ms.
}
\label{lambda_free_Dt}
\end{figure}

In order to measure the fringe wave vector $K_f$, we further integrate the 2D density profiles along the $z$ axis, and then we extract $K_f$ from the corresponding FT, see Fig.~\ref{lambda_free_Dt}, as a function of the evolution time $\Delta t$ (red points). In this figure we also show the result of GP simulations (continuous blue line), which nicely match the experimental data.

In the simulations, the condensate is described by a wave function $\psi(\bm{r},t)$ that evolves according to the GP equation \cite{dalfovo1999}
\begin{equation}
i\hbar\partial_{t}\psi = \left[-\frac{\hbar^{2}}{2m}\nabla^{2} + U(\bm{r},t) + g|\psi|^{2} \right]\psi,
\label{eq:GPE}
\end{equation}
with $g=4\pi\hbar^{2}a/m$ and $a=99a_0$ the $s$-wave scattering length. Here $U(\bm{r},t)$ represents the Bragg potential, with the same sequence as in the experiment \footnote{We first solve the stationary GP equation in the presence of a harmonic trap (as in the experiment), by means of a conjugate gradient algorithm \cite{press2007,modugno2003}. The time-dependent GP equation is solved by using the FFT split-step method discussed in \cite{jackson1998}. The typical size of the numerical grid is $96\mu$m$\times96\mu$m$\times48\mu$m, with $1024\times128\times64$ points.
The simulations are performed in the reference frame of the initial condensate (that is, by considering a moving optical lattice). In order to simplify the calculations, at $t_{TOF}=6$~ms we manually remove the momentum components that do not contribute to the interferograms in ports A and B, namely those with momentum components with $|k|>1.33k_L$. This cut in momentum space permits one to reduce the size of the numerical box. Then, thanks to the fact that the system becomes almost noninteracting already at $t_{TOF}=8$~ms, the subsequent expansion dynamics up to $t_{TOF}=33$~ms is via a \textit{free} expansion. The latter amounts to a single multiplication in Fourier space.}.

Regarding the effect of interactions on the interference pattern, it is instructive to compare the above results with the expected value of the fringe wave vector for the case of two condensates which interfere in the absence of mutual interactions, $K_{f,0}$:
in this case, since the interfering wave functions have no relative velocity, we have
(see e.g. Refs. \cite{simsarian2000,Fort2001} and the discussion in Sec.~\ref{sec:model})
\begin{equation}
K_{f,0}=\frac{m}{\hbar}\frac{\dot{\lambda}_x(t_{TOF})}{\lambda_x(t_{TOF})}\delta x
\label{lambda_teo}
\end{equation}
where $\delta{x}\equiv x_2-x_1= (2\hbar k_L/m) \Delta t$ is the separation between the condensate centers of mass,
$\lambda_{\nu}$ ($\nu=x,y,z$) are the dimensionless scaling parameters governing the expansion of the condensate in the Thomas-Fermi regime \cite{castin1996}, and $t_{TOF}$ represents the total expansion time. This result corresponds to the dashed blue curve in Fig.~\ref{lambda_free_Dt}, which displays a significant deviation with respect to the full GP results and the experimental data. This is not surprising, as here the condensates are indeed interacting.

Then, in order to account for the effect of the mutual repulsion, we consider an approach that has been often used in the literature, namely we assume that the phase of the two condensates is modified by an additional velocity term produced by an effective force due to the mean-field interaction of the two condensates \cite{simsarian2000,benton2011,guan2020}. According to the discussion in \cite{guan2020} [see their Eq. (34)], the effective force (that here acts along one of the strongly confined directions of the trap) can be written as
\begin{equation}
F_x(t)=\pm \frac{m \omega_x^2 |\delta x(t)|}{\lambda_x(t)^3\lambda_y(t)\lambda_z(t)}
\label{eq:Fx}
\end{equation}
where $\delta x(t)$ is the time-dependent distance between the centers of mass of the two expanding condensates and the sign $\pm$ refers to the right and left condensate, respectively \footnote{Following the discussion in Ref. \cite{guan2020}, here we include an extra factor of $2$ in the force owing to the fact that the two condensates are in the same internal state. The force is then normalized by a factor of $1/2$ each time a $\pi/2$ pulse is applied, in order to account for the halving of the population of each component; see again Ref. \cite{guan2020}.}.
Though not explicitly mentioned in Ref. \cite{guan2020}, in the derivation of Eq. (\ref{eq:Fx}) there is the implicit assumption that the two condensates substantially overlap. In our present setup, this condition restricts its limits of applicability to $\Delta{t}\lesssim1$ ms.
The above expression of $F_{x}(t)$ is then used to compute the position and velocity of the center of mass of the two condensates at $t=t_{TOF}$. Finally, the fringe wave vector is evaluated through the Eq.~(\ref{eq:kf}) discussed in Appendix~\ref{appendix:B}, and it is shown in Fig.~\ref{lambda_free_Dt} as a dotted blue line. Remarkably, it turns out that this effective approach
overestimates the effect of interactions and it produces a wave vector smaller than observed.
We shall see in Sec. \ref{sec:model} that the failure of this approach resides in the assumption of ``rigidity'' of the condensate phase, that in general is not justified in the presence of local interactions, as they are likely to produce local variation of the phase.

\begin{figure}[ht!]
\begin{center}
\includegraphics[width=\columnwidth]{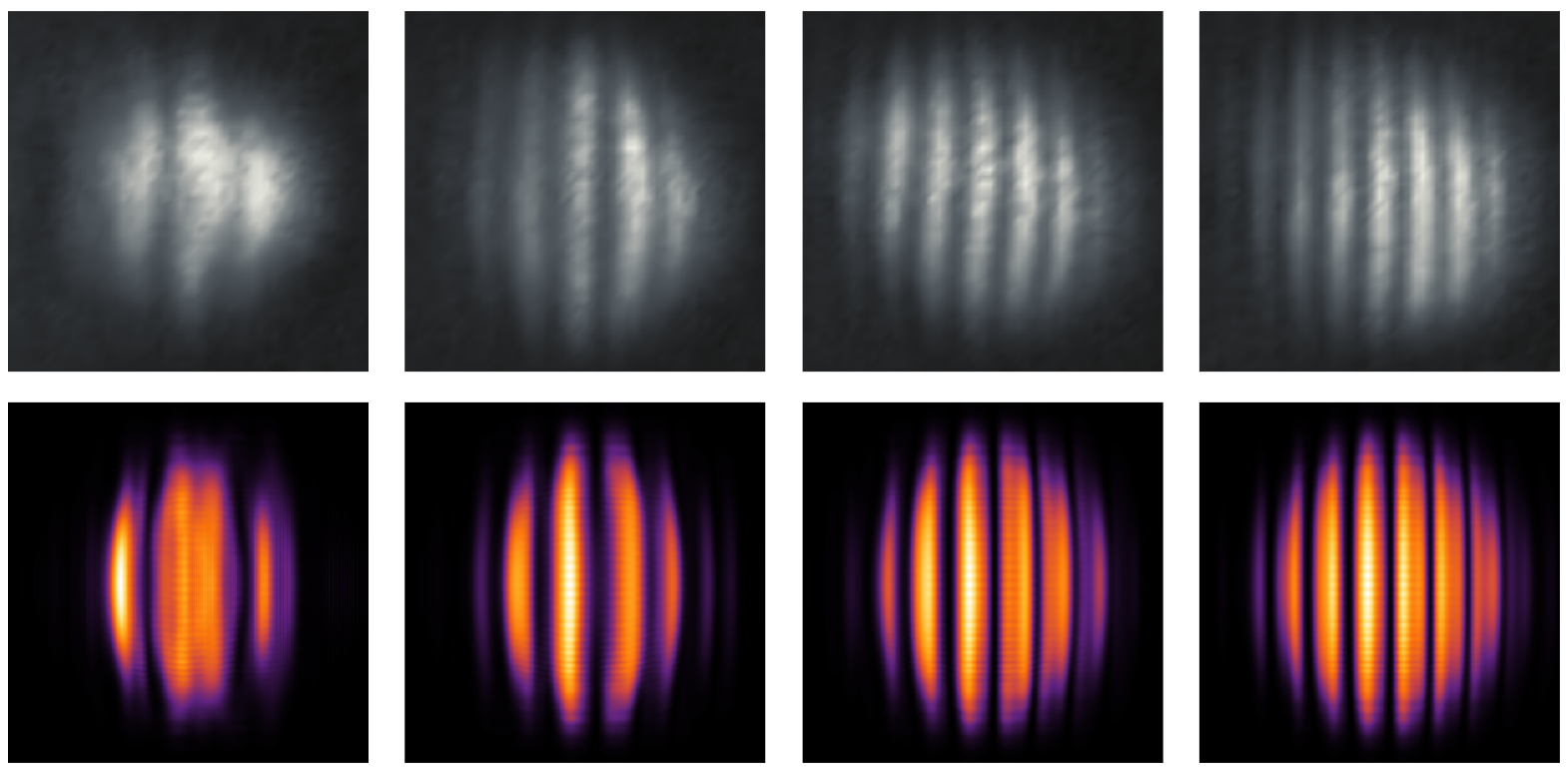}
\end{center}
\caption{Column density in the $xz$ plane measured at port B of the interferometer,
at $t=t_{TOF}$.
Here we vary $t_{delay}$ (0.5~ms, 1.5~ms, 2.5~ms, and 3.5~ms, from left to right) while $\Delta t=1$~ms is fixed.
First row: experimental data with a mean atom number $N=5\cdot 10^5$ and harmonic  frequencies of the initial trap
$(\omega_x,\omega_y,\omega_z)\simeq2\pi~\times~(55, 15, 65)$~Hz. Second row:
GP simulations. The box size is 192 $\mu$m.}
\label{free_space_delay}
\end{figure}

In a second series of measurements, we kept constant the time between the two Bragg pulses, $\Delta t=1$~ms, and we varied $t_{delay}$ from 0.5~ms to 4.0~ms, which amounted to varying the BEC density entering the interferometer with
$\delta x$ almost constant (except for small effects of the interactions). Samples of the column density patterns measured at port B are shown in Fig.~\ref{free_space_delay}, along with the results of the GP simulation for the same sets of parameters.

\begin{figure}[htb]
\begin{center}
\includegraphics[width= 0.9\columnwidth]{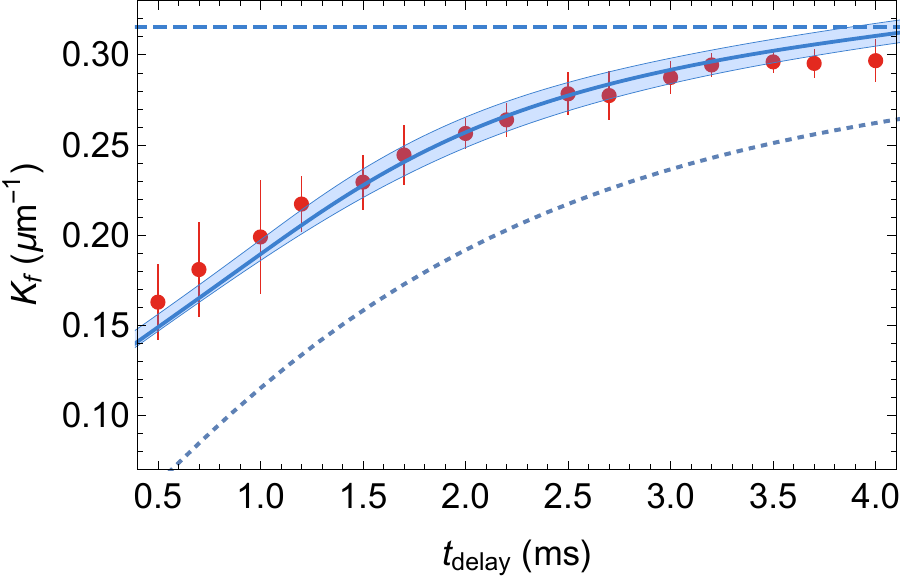}
\end{center}
\caption{Fringe wave vector $K_f$
as a function of $t_{delay}$, for fixed $\Delta t=1$~ms.
The red points correspond to the experimental data, the dashed blue line is the theoretical prediction of Eq.~(\ref{lambda_teo}), and the continuous blue line represents the results of the GP simulations (with a band representing a 50\% uncertainty on $N$). The dotted blue line is calculated correcting the center-of-mass motion of the two condensates using the effective force in Eq. (\ref{eq:Fx}), as in Fig. \ref{lambda_free_Dt}.
}
\label{lambda_free_delay}
\end{figure}

The corresponding values of the fringe wave vector $K_f$ are shown in Fig.~\ref{lambda_free_delay} (red points), along with the prediction of Eq.~(\ref{lambda_teo}) (dashed blue line), the results obtained using the effective force in Eq.~(\ref{eq:Fx}) (dotted blue line) and the result of full GP simulation (continuous blue line). As one may naively expect, this figure shows that the effects of interactions are predominant at short times where atomic densities are larger, whereas for longer times, since the atomic densities are lower, the fringe spacing approaches the prediction for the noninteracting case.
Again we see that the center-of-mass motion induced by the force $F_x$ implies a wave vector variation (blue dotted line in Fig.~\ref{lambda_free_delay}) larger than the experimental data and the GP results \footnote{Notice that in both Figs. \ref{lambda_free_Dt} and \ref{lambda_free_delay} we have considered only the mutual effect of the two wave packets belonging to the same port of the interferometer. We have verified that by including all the four wave packets (produced after the second Bragg pulse) in the description, this results in additional forces that make the wave vector $K_{f}$ decrease even further. Similarly, if one corrects the Castin-Dum equations \cite{castin1996} for the scaling parameters $\lambda_{\nu}(t)$ in order to account for the splitting of the total number of atoms across the two interferometer ports, this also lowers the dotted line.}.

\begin{figure}[t!]
\begin{center}
\includegraphics[width= 1\columnwidth]{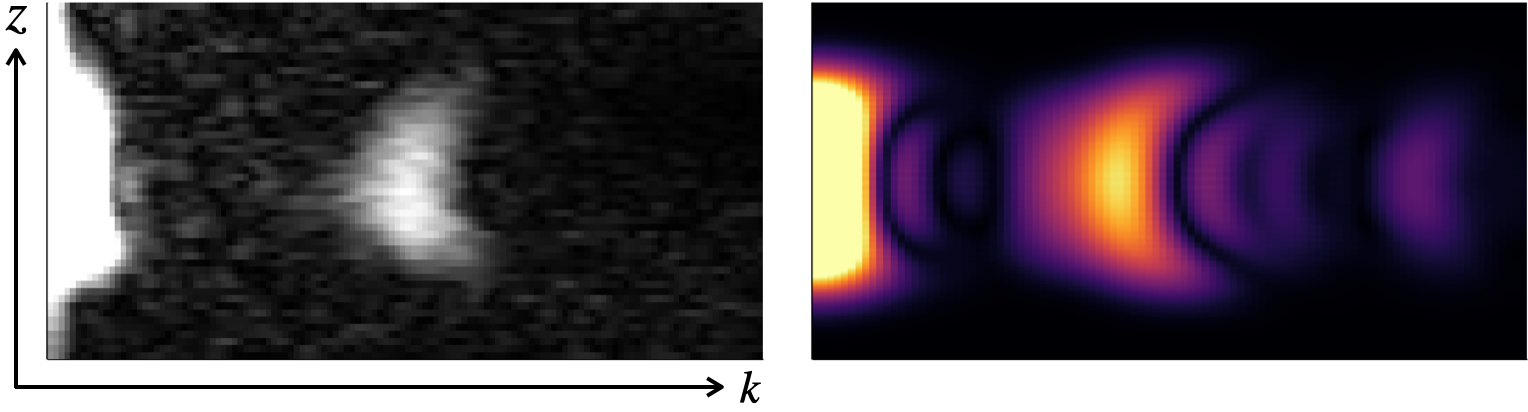}
\end{center}
\caption{FT row by row of the column  density distribution in the $xz$ plane for $\Delta t=1.6\,$ms and $t_{delay}=1\,$ms: experiment (left) and GP simulation (right). The box size is $0.82\,\mu$m$^{-1}$ (wave vector space, horizontal) and $192\, \mu$m (coordinate space, vertical).}
\label{fig:slices}
\end{figure}

Till now, we extracted $K_f$ by analyzing the density profiles integrated along the $y$ and $z$ directions, thus averaging over different densities' regions of the atomic cloud.
The effect of interactions can be highlighted in a single measurement if we evaluate the interference fringe wave vector for different sections of the atomic density distribution along the $z$ axis. In Fig. \ref{fig:slices} we show the FT row by row of a single image, i.e., for varying $z$ coordinates: in the outer regions of low (column) density the fringes are thinner and,
correspondingly, the peak of the Fourier transform moves towards larger wave vectors, in agreement with the trend displayed in Fig. \ref{lambda_free_delay}. Thus the interferograms display peculiarly curved fringes that have been previously observed \cite{simsarian2000} and studied \cite{rohrl1997, liu2000}. Interestingly, the semianalytical models assuming the phase as rigidly determined by the center-of-mass motion of the condensates predict the same fringe wave vector, independent of the density variation along the $z$ axis (see Appendix~\ref{appendix:expansion}).

Our observations clearly show that the mutual repulsive interactions of two interfering condensates
produce quantitative modifications of their interferogram. Though similar effects have already been pointed out in previous
experiments \cite{simsarian2000,guan2020}, and have been addressed by
various theoretical works \cite{rohrl1997,jamison2011,benton2011,guan2020}, still a clear explanation is lacking. In particular,
the assumption that interaction effects can be accounted for by an effective force rigidly altering the phase of the expanding condensates through their center-of-mass motion seems not fully justified in our case. Then, in order to make a further step and clarify this matter, in the following we consider a simplified scenario which captures the essential features of the experiment and permits a thorough theoretical analysis of the effects of the mutual repulsion on the phase of two expanding condensates.

\section{Theoretical analysis of the phase}
\label{sec:model}

\begin{figure}[ht!]
\centerline{\includegraphics[width=0.9\columnwidth]{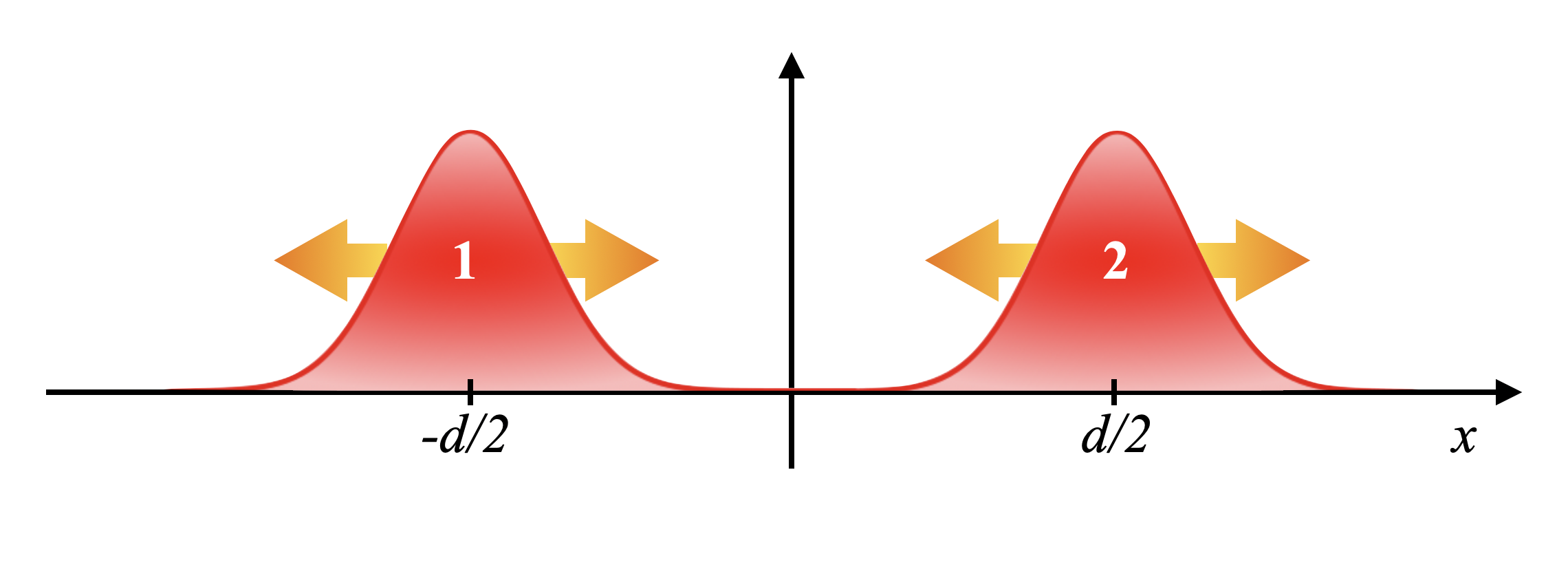}}
\caption{Sketch of the system considered here: two condensates with the same density distribution, initially placed at distance $d$, are let expand freely until they overlap and interfere (when projected onto the same quantum state).}
\label{fig:sketch}
\end{figure}

Two condensates with identical density distribution, are placed at distance $d$ along the $x$ direction, as shown in Fig. \ref{fig:sketch}. At time $t=0$ the two condensates are released from the trapping potential and let expand.
Each condensate, consisting of $N/2$ atoms, is prepared in the ground state of an axially symmetric harmonic trap, $\omega_{x}=\omega_{z} > \omega_{y}$. As for the interaction, for simplicity we restrict the analysis to the symmetric case $g_{11}=g_{12}=g_{22}\equiv g$ (this is an excellent approximation for experiments with $^{87}$Rb).

For clarity, we start by considering the case in which
the atoms in the two condensates occupy \textit{different internal quantum states} (that we indicate as $|\!\uparrow\rangle$ and $|\!\downarrow\rangle$), so that the corresponding wave functions evolve according to two coupled Gross-Pitaevskii (GP) equations.
This choice has the advantage that it allows one to clearly identify the phase of each condensate, at any time. Then, we shall extend the discussion to the experimental situation, in which the two condensates are formed by atoms in the \textit{same internal quantum state}, and, despite being at a distance $d$ apart, are described by a single coherent wave function. These two scenarios are analogous to those considered in \cite{guan2020}, with the difference that here we do not consider two different momentum states, but rather two wave packets that are spatially separated, initially.
In the first case [that we will refer to as (A)], the two condensates are in different internal states, when they overlap an instantaneous $\pi/2$ pulse is (ideally) applied to mix $|\!\!\downarrow\rangle$ and $|\!\!\uparrow\rangle$, and the density in one internal state becomes
$|\psi_{1}(\bm{r},t)+\psi_{2}(\bm{r},t)|^{2}$, with the mixed terms producing the interference pattern. In the second case [referred to as (B)], with all atoms in the same internal state, the same interference term appears naturally from different pieces of the total wave function $\psi(\bm{r},t)$, as it will be clear later on.

In the following, we shall discuss how the phase of each condensate is affected by the mutual interaction, and how this determines the wave vector of the interference pattern. Since the overlap between the two condensates occurs along the $x$ direction, we will focus on the behavior of the phase along the $x$ axis, namely for $y=z=0$. In particular, we want to provide a quantitative answer to the following question: \textit{is it possible to describe the modification of condensate phase in terms of the effective force that determines the center-of-mass motion of the two interacting condensates?}

In order to do so, in the rest of this section we analyze the phase resulting from the GP evolution, and we compare it with the expected behavior (in dimensionless units, see Appendix~\ref{appendix:expansion})
\begin{equation}
\Phi(x,t)=\frac12 \frac{\dot\lambda(t)}{\lambda(t)}x^{2} + \left[\dot\alpha(t)-\frac{\dot\lambda(t)}{\lambda(t)}\alpha(t)\right]x.
\label{eq:phase_acc_motion}
\end{equation}
 We recall that the above expression has been obtained assuming that the mutual repulsion between the two condensates can be described in terms of an effective force $F(t)$, that determines the condensates center-of-mass motion.
In this framework, $\alpha(t)$ represents the position of the condensate center of mass at time $t$ and $\dot\alpha(t)$ the corresponding velocity.

\subsection{Different internal states}
\begin{figure}[b!]
\centerline{\includegraphics[width=0.9\columnwidth]{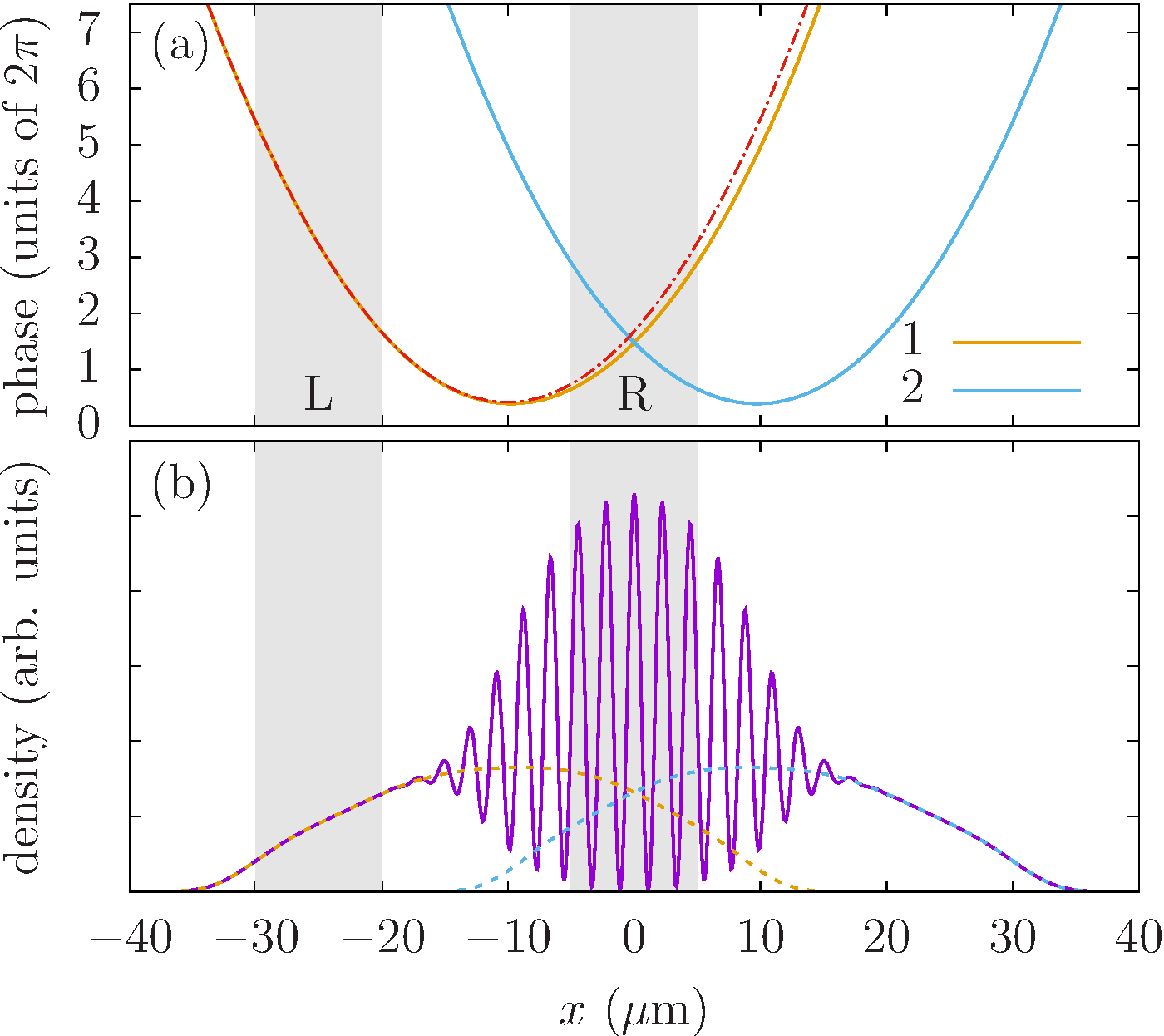}}
\caption{Interference pattern for case A at $t=8$ ms. (a) Plot of the phase of the two condensates (solid lines), and of $\Phi_{L}^{fit}$ (dotted-dashed line); (b) the density $n_{\uparrow}(x,0,0)$ of the $|\!\!\uparrow\rangle$ state after a $\pi/2$ pulse (solid line), along with the density distribution $|\psi_{i}(x,0,0)|^{2}$ of the two condensates just before the pulse (dashed lines). The shaded areas represent the two regions of the fit (see text).}
\label{fig:density-phase}
\end{figure}

We start with the case of two condensates in different internal states, described by the wave functions $\psi_{1,2}(\bm{r},t)$, which obey the following  GP equations \cite{leggett2001}:
\begin{equation}
\begin{cases}i\hbar \partial_{t}\psi_{1} =
\displaystyle\left[-\frac{\hbar^{2}}{2m}\nabla^{2} + g|\psi_{1}|^{2} + g|\psi_{2}|^{2}\right]\psi_{1}
\\
i\hbar \partial_{t}\psi_{2} =
\displaystyle\left[-\frac{\hbar^{2}}{2m}\nabla^{2} + g|\psi_{2}|^{2} + g|\psi_{1}|^{2}\right]\psi_{2}.
\end{cases}
\label{eq:2GPE}
\end{equation}
The two condensates are initially displaced by a distance $d=20$ $\mu$m, and expand for a variable time up to $10$ ms.
In Fig. \ref{fig:density-phase} we show the phase of the two components and the density of the $|\!\!\uparrow\rangle$ state after a recombining $\pi/2$ pulse, at $t=8$ ms, when the two condensates are substantially overlapped.
Since the two phases $\Phi_{i}(x,t)$ are expected to be quadratic functions of the coordinate $x$ [see Eq.~(\ref{eq:phase_acc_motion})], and indeed Fig. \ref{fig:density-phase}(a) shows that they have an \textit{almost} parabolic shape \footnote{In order to represent the phase as a smooth curve, we remove artificial jumps produced by the fact that the phase can only be determined modulo $2\pi$. This permits onto ``unwrap'' the values obtained from the numerics, in order to keep a (piecewise) monotonic behavior.}, it is natural to fit the numerical data with the following expression
\begin{equation}
\label{eq:phase4}
\Phi_{\ell}^{fit}(x,t)=\frac12a_{\ell}(t)x^{2} + b_{\ell}(t)x + c_{\ell}(t),
\end{equation}
where $a_{\ell},b_{\ell},c_{\ell}$ are fitting parameters. The index $\ell=L,R$, not to be confused with $i=1,2$, is introduced because we are going to perform two independent fits of the phase -- one in the interference region where the two condensate overlap and the other in the outer portion of the condensate. In particular, here we are going to refer to the phase $\Phi_{1}(x,t)$ of the leftmost condensate, so that the outer and inner regions correspond to L and R, respectively (see the shaded areas in Fig. \ref{fig:density-phase}).

\begin{figure}[t!]
\centerline{\includegraphics[width=0.9\columnwidth]{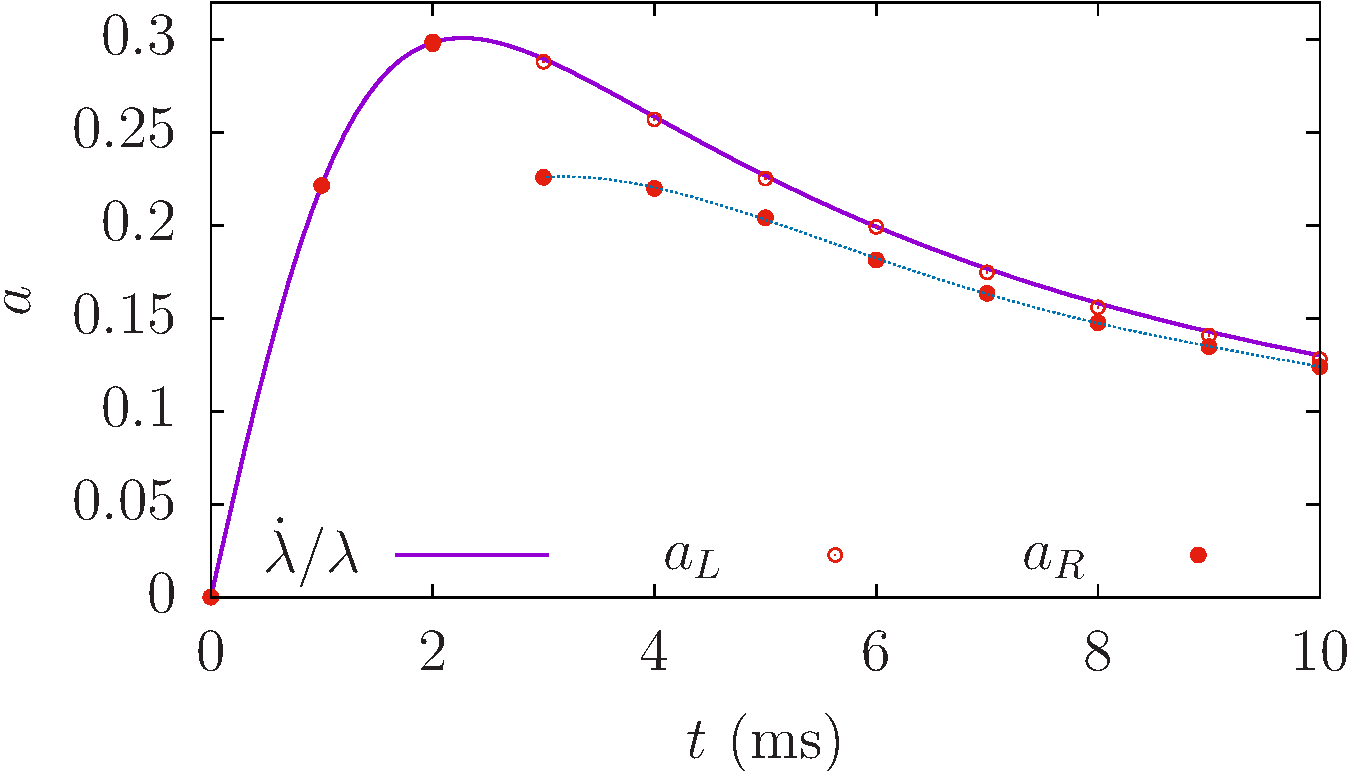}}
\caption{Fitted values of $a_{L,R}(t)$ compared with the expected behavior ${\dot\lambda(t)}/{\lambda(t)}=t/(1+t^2),$ both expressed in dimensionless units. The dotted line is a guide to the eye.}
\label{fig:afit}
\end{figure}

Let us now analyze the result of the fit from a quantitative viewpoint. In Fig. \ref{fig:afit} we compare the fitted values of $a_{L,R}(t)$ with the prediction of Eq. (\ref{eq:phase_acc_motion}), assuming the approximate expression $\lambda(t)=\sqrt{1+t^{2}}$, namely $\dot\lambda(t)/\lambda(t)=t/(1+t^2)$. This figure shows that $a_{L}$ nicely follows the expected behavior, whereas $a_{R}$ displays a significant deviation at $t\simeq3$~ms when, during the expansion, the two condensates start to overlap.
This is a first hint of a local effect of the interaction between the two condensates, which cannot be accounted for by a global modification of the phase.

\begin{figure}[t!]
\centerline{\includegraphics[width=0.9\columnwidth]{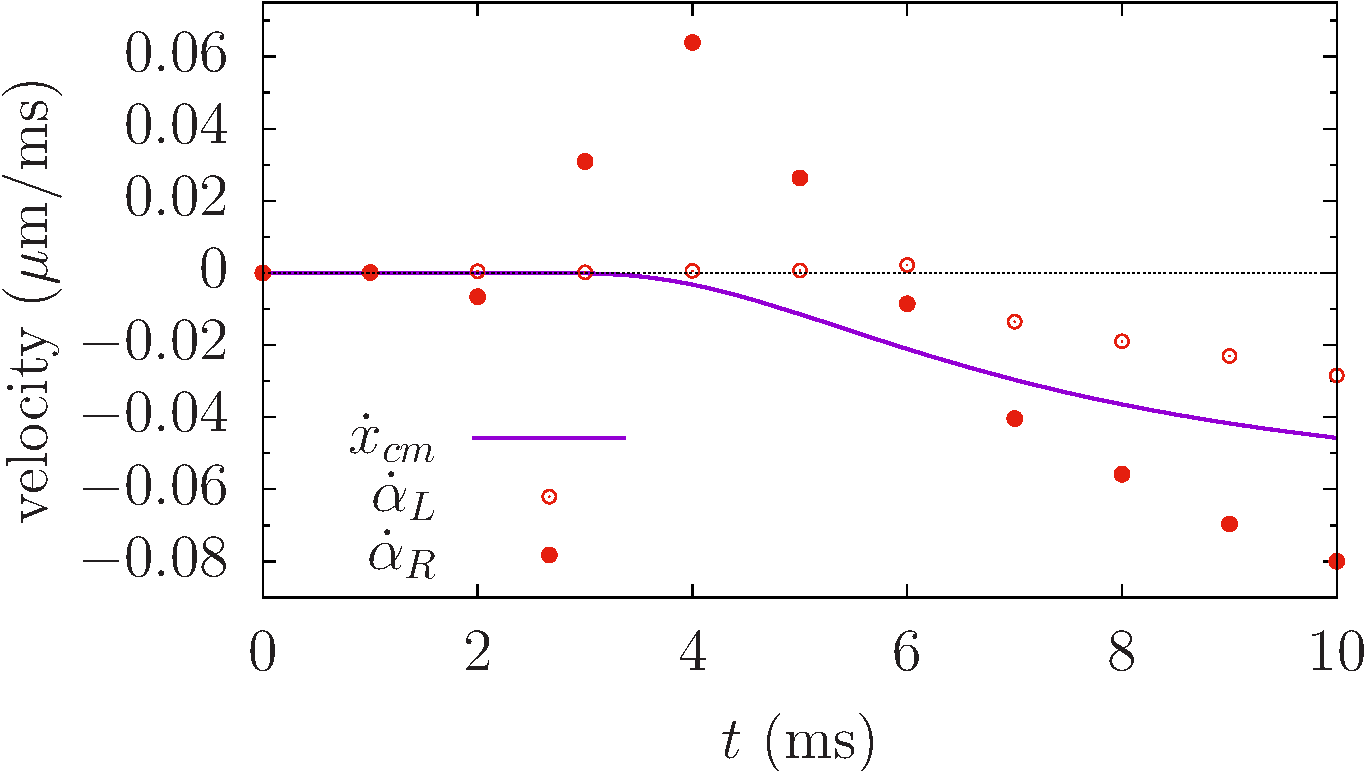}}
\caption{Values of $\dot\alpha_{\ell}(t)$ ($\ell=L,R$) extracted from the fit of the phase [see Eq. (\ref{eq:alphadot})] (red circles) compared with the center-of-mass velocity $\dot{x}_{cm}(t)$ (solid line). This figure shows that the linear term of the phase is not related to the center-of-mass velocity.}
\label{fig:velocity}
\end{figure}

We now turn to the linear term.
According to Eq. (\ref{eq:phase_acc_motion}), the expected value of $b_\ell(t)$ is
\begin{equation}
b_{\ell}(t)=\dot\alpha(t)-\frac{\dot\lambda(t)}{\lambda(t)}\alpha(t).
\label{eq:for-b}
\end{equation}
In order to verify this relation, we proceed as follows. First, we indicate with $x_{cm}(t)\equiv\int x|\psi_{1}(\bm{r})|^{2}d\bm{r}$ the $x$ coordinate of the first condensate center of mass, obtained from the numerical simulations, and with $\dot{x}_{cm}(t)$ the corresponding velocity. Then, by replacing $\alpha(t)$ with $x_{cm}(t)$  and   $\dot\lambda(t)/\lambda(t)$ with $a_{\ell}(t)$ in Eq. (\ref{eq:for-b}), we define
\begin{equation}
\dot\alpha_{\ell}(t)=b_{\ell}(t)+a_{\ell}(t)x_{cm}(t),
\label{eq:alphadot}
\end{equation}
that we compare with $\dot{x}_{cm}(t)$, in Fig. \ref{fig:velocity}.
Again, this figure reveals that the value of $\dot\alpha_{\ell}(t)$ in Eq. (\ref{fig:velocity}), resulting from the fit of the condensate phase, presents a substantial deviation from the center-of-mass velocity obtained from the numerical solution of the GP equations.

\begin{figure}[h!]
\centerline{\includegraphics[width=0.9\columnwidth]{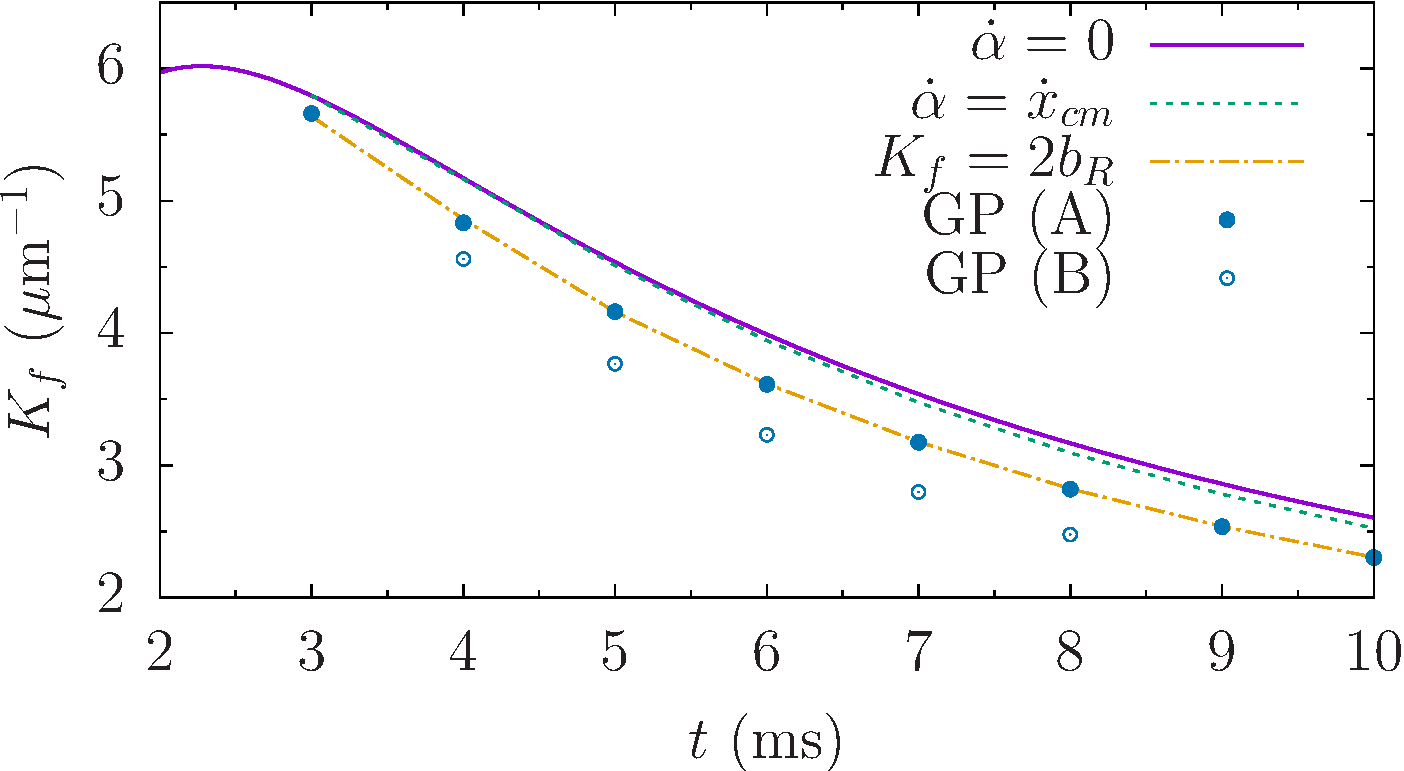}}
\caption{Fringe wave vector $K_{f}$ as a function of time.
The values extracted from a sinusoidal fit of the density modulations [see, e.g., Fig. \ref{fig:density-phase}(b)] as obtained from the GP simulations (for cases A, B) are compared to the formula in Eq. (\ref{eq:kf}), for different settings.
The continuous line corresponds to the case in which the two condensates do not interact ($\dot\alpha=0$), the dashed line to $\dot\alpha=\dot{x}_{cm}$, and the dotted-dashed line obtained from the fitted value of $b_{R}$, $K_{f}=2b_{R}$. The excellent agreement between the latter and the GP values (for the present case, A) provides a consistency check between the fits of the wave-function phase and of the fringes, which are \textit{independent}.
}
\label{fig:lambda}
\end{figure}

The above result indicates that in general it is not justified to describe the phase simply in terms of the effective force that determines the center-of-mass motion of the two interacting condensates, as it is usually assumed in the literature \cite{simsarian2000,guan2020}. This fact is particularly relevant for experiments that use condensate for interferometric purposes. Indeed, the deviation from the simple behavior in Eq. (\ref{eq:phase_acc_motion}) leads to \textit{measurable effects} in the wavelength of the interference pattern. A comparison between the analytical expression in Eq. (\ref{eq:deltaphi}) and the value of the fringe wave vector $K_{f}$ extracted from a sinusoidal fit of the fringes \footnote{Here we employ a direct fit in coordinate space instead of using the FT approach because we are interested in measuring the \textit{local} effect of the phase.}, at different evolution times, is shown in Fig. \ref{fig:lambda}.
The explanation for this behavior is the following: the condensate's phase, and therefore the fringe wave vector,
are locally affected by the interactions, not through the center-of-mass velocity.
\subsection{Same quantum state}
In this case the whole system can be described by a single wave function $\psi$ that evolves according to
the same GP equation as in Eq. (\ref{eq:GPE}), with $U\equiv0$.
In order to proceed with the same analysis as in the previous case, we can conveniently split the wave function in two components, $\psi=\psi_{1}+\psi_{2}$, that we associate to the two initial condensates; see Fig. \ref{fig:sketch}. Then, it is straightforward to prove that
Eq. (\ref{eq:GPE}) (with $U\equiv0$) is formally equivalent to the following two coupled GP equations
\begin{equation}
\label{eq:2GPE-new}
\!\!\!
\begin{cases}
i\hbar \partial_{t}\psi_{1}\! =\!\!
\displaystyle\left[-\frac{\hbar^{2}}{2m}\nabla^{2} \!+ g\sum_{i=1}^{2}|\psi_{i}|^{2} \!\!+\! 2g\textrm{Re}(\psi_{1}^{*}\psi_{2})\!\right]\!\psi_{1}
\\
i\hbar \partial_{t}\psi_{2}\! =\!\!
\displaystyle\left[-\frac{\hbar^{2}}{2m}\nabla^{2} \!+ g\sum_{i=1}^{2}|\psi_{i}|^{2} \!\!+\! 2g\textrm{Re}(\psi_{1}^{*}\psi_{2})\!\right]\!\psi_{2}.
\end{cases}\!\!\!\!\!\!\!\!
\end{equation}
that differs from Eq. (\ref{eq:2GPE}) owing to the presence of the term proportional to $\textrm{Re}(\psi_{1}^{*}\psi_{2})$. This manifestly shows that here the role of interactions can be more complex, with respect to the previous case of two different quantum states. We remark that the choice of the splitting is not unique -- the mean-field term $|\psi_{1}+\psi_{2}|^{2}(\psi_{1}+\psi_{2})$ could be split differently between the two equations, see, e.g., Ref. \cite{guan2020} -- and somewhat arbitrary, as the two components are distinguishable only initially, when they are spatially separated. Nevertheless, it produces a mean-field potential that is defined real, and it is symmetric under the permutation $1\leftrightarrow2$. Such a splitting is useful for identifying two components which eventually interfere, and to keep track of the corresponding phases.

\begin{figure}[t!]
\centerline{\includegraphics[width=0.9
\columnwidth]{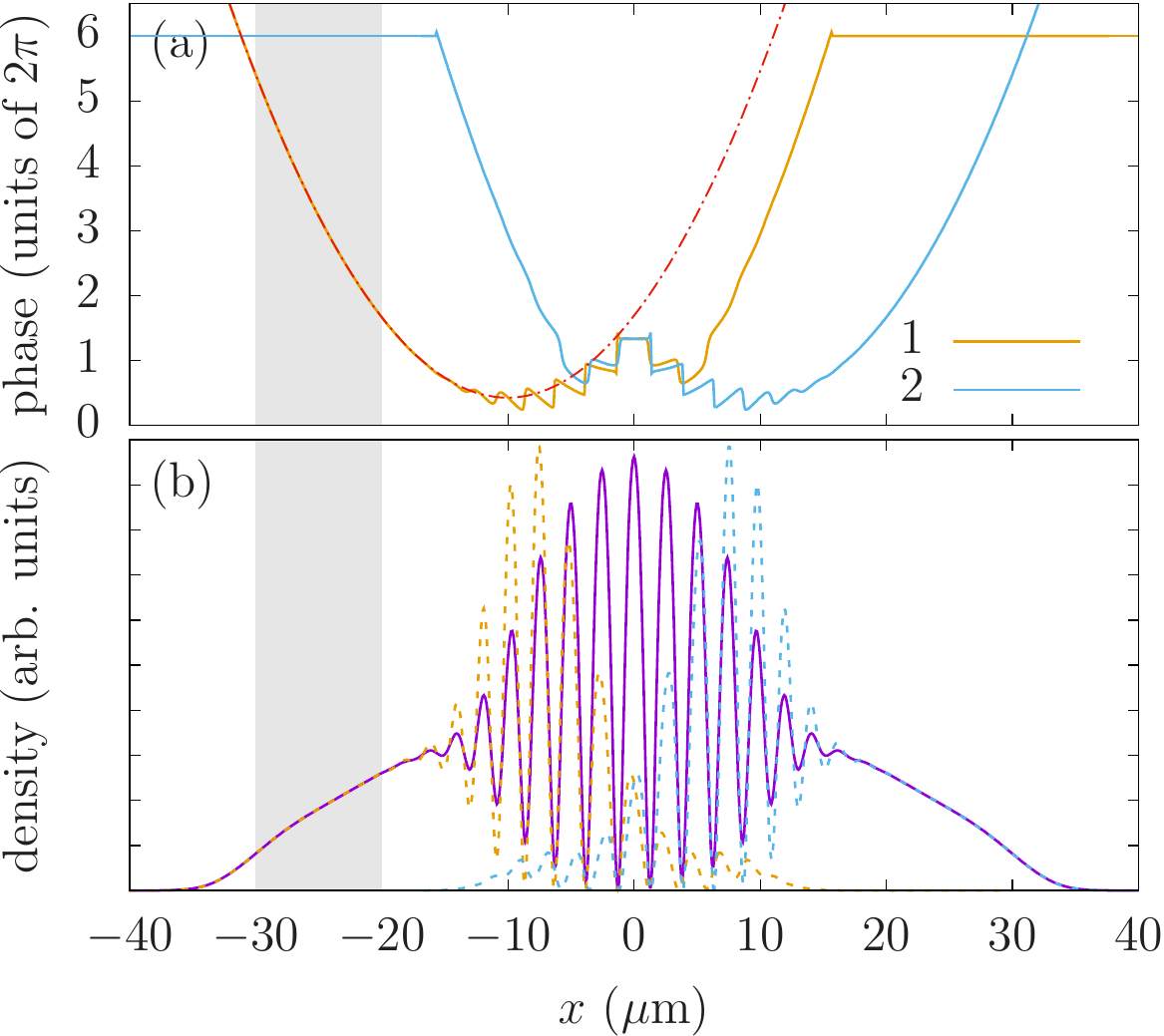}}
\caption{Interference pattern for case B at $t=8$ ms. (a) Plot of the phase of $\psi_{1}$ and $\psi_{2}$ (solid lines), and of $\Phi^{fit}$ (dotted-dashed line); (b) total density $|\psi_{1}+\psi_{2}|^{2}$ (solid line), along with $|\psi_{1}|^{2}$ and $|\psi_{2}|^{2}$. All quantities are plotted as a function of $x$, for $y=z=0$. The shaded area represents the region where we fit the phase of $\psi_{1}$ (see text).}
\label{fig:density-phase-B}
\end{figure}

Then, we proceed as in the previous case. The two condensates are initially displaced by a distance $d=20$ $\mu$m, and are let expand for a variable time, up to $10$ ms. The total density and the phase of the two components are shown in Fig. \ref{fig:density-phase-B}, at $t=8$ ms.
In this case the phases $\Phi_{i}(x,t)$ display a clean parabolic shape only outside the interference region, so that we restrict
the quadratic fit, see Eq. (\ref{eq:phase4}) only to the leftmost region -- the shaded area in Fig. \ref{fig:density-phase-B} --
and remove the index $\ell$. We find that the coefficient of the quadratic term, $a(t)$, behaves like $a_{L}$ in Fig. \ref{fig:afit}, nicely following the expected behavior ${\dot\lambda(t)}/{\lambda(t)}=t/(1+t^{2})$. Needless to say, this has no influence on the fringe spacing, as the latter is affected only by the phase in the overlap region. Notice that here the phase is characterized by ``jumps'' in correspondence of the density minima, associated to the presence of (quasi) nodes. The amplitude of the jumps is always less than or equal to $\pi$: they range from low amplitude knees that bend smoothly around finite size density minima to sudden jumps of amplitude $\pi$ in correspondence of density nodes, where the wave function changes sign.

\begin{figure}[t!]
\centerline{\includegraphics[width=0.9\columnwidth]{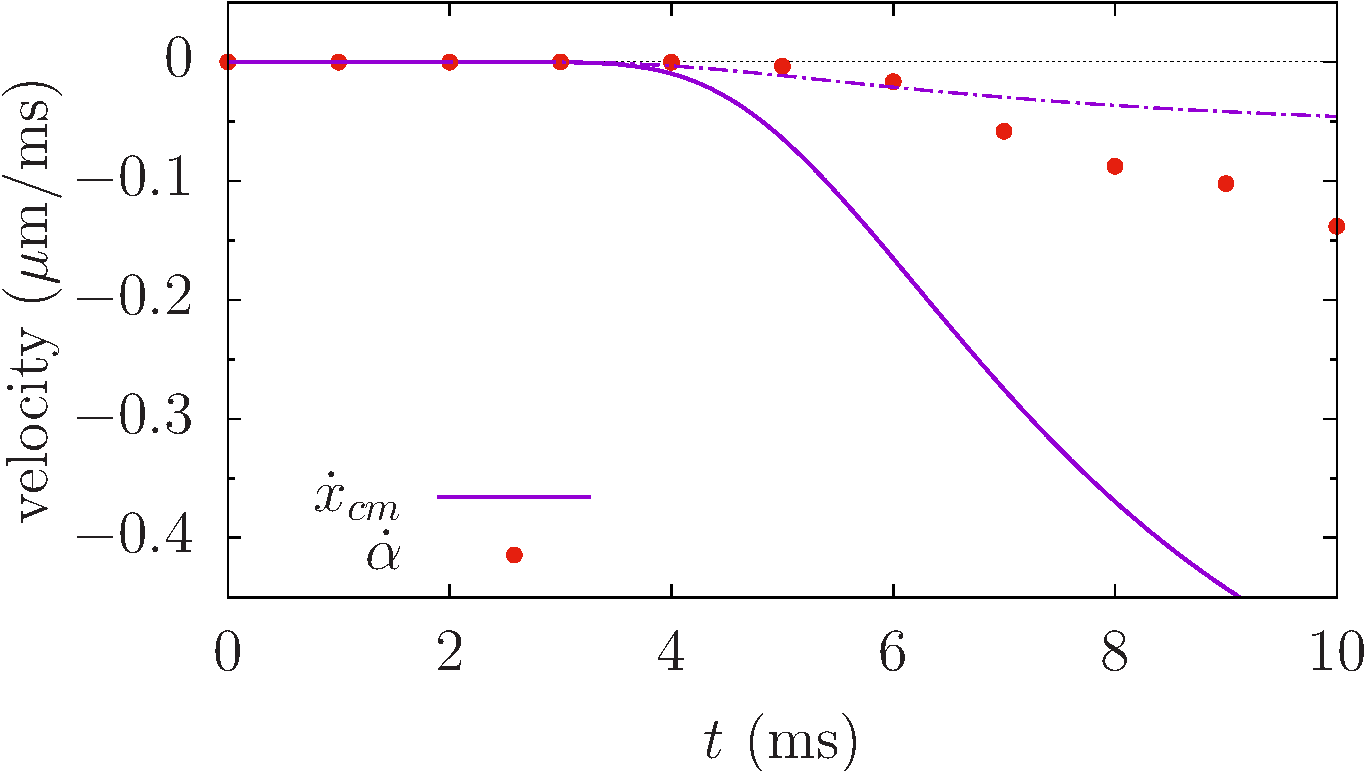}}
\caption{Values of $\dot\alpha(t)$ extracted from the fit of the phase [see Eq. (\ref{eq:alphadot})] (red circles) compared with the center-of-mass velocity $\dot{x}_{cm}(t)$ (solid line). As a reference we also show the center-of-mass velocity of the previous case (dotted-dashed line); see Fig. \ref{fig:velocity}.}
\label{fig:velocity-B}
\end{figure}
As for the velocity term, the combination of Eq. (\ref{eq:alphadot}) with the result of the fit is shown in Fig. \ref{fig:velocity-B}, along with the center-of-mass velocity $\dot{x}_{cm}(t)$ (solid line) and the corresponding curve for the previous case in Fig. \ref{fig:velocity}.
From this figure we evince that: (i) the value of $\dot\alpha_{\ell}(t)$ inferred from the fit of the phase displays a larger deviation from the actual center of mass, with respect to the case treated in the previous section, and (ii) the effect of the mutual repulsion on the center-of-mass motion of the two components is stronger than in the previous case. The fact that in the present case B (condensates in the same quantum state) interactions play a more relevant role with respect to the previous case A (different quantum states), is also evident from the values of the fringe wave vector shown in Fig. \ref{fig:lambda} (empty circles). Indeed, in the present case the deviation from the prediction for the noninteracting case is substantially larger than in case A.

\section{Summary and Outlook}
\label{sec:SO}

We have presented a quantitative investigation of the interference produced by two expanding condensates where mutual interactions play a substantial role. The wave vector of the
interferogram fringes has been measured varying the timing of an interferometer consisting in two $\pi/2$ Bragg pulses.
The experimental data have been compared with two semianalytical models obtained either by neglecting altogether the mutual interactions or by accounting for them through an effective force that globally alters the center-of-mass dynamics, and hence the condensates' phases, finding that this approach fails to quantitatively match the observations. We have also performed GP simulations which remarkably agree with the experiment. Motivated by these results, in order to clarify
the effect of mutual interactions on the phase of the interfering condensates, we have introduced a simple but instructive model of two initially separated condensates which interact and interfere during the expansion \cite{andrews1997}.
By analyzing their phase evolution we reach the main conclusion that we have to abandon the idea of a rigid phase, globally modified by the interactions.
Instead, we have shown that interactions affect the phase only \textit{locally}, in the region where the two interfering wave packets superimpose.

These results, which we believe have a general validity for a deeper understanding of matter wave interference, open perspectives for interferometric applications of interacting Bose-Einstein condensates.

\acknowledgments
We acknowledge fruitful discussions with M. Fattori, L. Masi, M. Prevedelli, R. Corgier, and A. Smerzi and we thank M. Inguscio for continuous support.
This work was supported by the Spanish Ministry of Science, Innovation and Universities and the European Regional Development Fund FEDER through Grant No. PGC2018-101355-B-I00 (MCIU/AEI/FEDER, UE), by the Basque Government through Grant No. IT986-16, by the European Commission through FET Flagship on Quantum Technologies-Qombs Project (Grant No. 820419) and by Fondazione Cassa di Risparmio Firenze through project ``SUPERACI-Superfluid Atomic Circuits."

\appendix

\section{Expansion of a single wave packet in the presence of an external force}
\label{appendix:expansion}

Here, we recall how a single condensate expands in the presence of an external force along the $x$ axis,
generated by the linear potential $U(x)=-F(t)x$.
The condensate wave function $\psi(\bm{r},t)$ evolves according to the GP equation in Eq. (\ref{eq:GPE})
that admits solutions of the form
\begin{equation}
\psi(\bm{r},t) = \varphi(x-\alpha(t),y,z;t)e^{im\beta(t) x/\hbar}e^{-i\gamma(t)/\hbar}.
\label{eq:rescaled}
\end{equation}
 Then, by setting
\begin{equation}
\label{eq:alpha}
\dot\alpha=\beta,\quad \dot\beta = F/m,\quad \dot\gamma=2\beta^{2}/m,
\end{equation}
it is straightforward to get
\begin{equation}
i\hbar \partial_{t}\varphi(\bm{\xi},t) =
\left[-\frac{\hbar^{2}}{2m}\nabla_{\bm{\xi}}^{2} + g|\varphi(\bm{\xi},t)|^{2} \right]\varphi(\bm{\xi},t).
\label{eq:reducedGPE}
\end{equation}
where we have defined $\bm{\xi}\equiv \bm{r}-\alpha(t)\bm{e}_x$ and $\bm{e}_x$ is the unit vector in $x$ direction.
This equation, along with Eq. (\ref{eq:rescaled}), tells us that if $\varphi(x,t)$ is the wave function of a freely expanding condensate, the wave function of a condensate that expands in the presence of the force $\bm{F}=F(t)\bm{e}_{x}$ is obtained from the former as
\begin{equation}
{\psi(\bm{r},t) = \varphi(x-\alpha(t),y,z;t)e^{im\dot\alpha(t) x/\hbar}e^{-i\gamma(t)/\hbar}},
\label{eq:rescaled2}
\end{equation}
where $\alpha(t)$ is obtained by integrating Eqs. (\ref{eq:alpha}).

In the Thomas-Fermi regime relevant for the experiment, the expansion of each of the two condensates is characterized by an almost self-similar behavior \cite{dalfovo1999} described by the Castin-Dum scaling parameters \cite{castin1996,guan2020}. In particular, the $x$-dependent component $\phi(x,t)$ of the phase of a condensate initially in the ground state of a harmonic potential of frequency $\omega_{x}$, and centered in $x=0$, is described by the following simple analytical expression:
\begin{equation}
\phi(x,t)=\frac12\frac{x^{2}}{a_{ho}^{2}}\frac{1}{\omega_{x}}\frac{\dot\lambda(t)}{\lambda(t)}
\end{equation}
where $a_{ho}=\sqrt{\hbar/(m\omega_{x})}$ is the harmonic-oscillator length, and
$\lambda(t)\simeq \sqrt{1+\omega_{x}^{2}t^{2}} $ is the transverse scaling parameter for an elongated trap \cite{castin1996}.
In order to simplify the notations, it is convenient to use dimensionless variables, by introducing $a_{ho}$ as length scale and $\omega_{x}$ as time scale. Then, velocities are measured in units of $\sqrt{\hbar\omega_{x}/m}$. In the rest of this section all quantities are assumed to be dimensionless (this corresponds to setting $\hbar=1=m$ in the previous expressions).

Then, the overall phase of $\psi$ in Eq. (\ref{eq:rescaled2}) can be written as $\Phi(x,t) + \Phi'(y,z,t)$, where
\begin{equation}
\Phi(x,t)=\frac12 \frac{\dot\lambda(t)}{\lambda(t)}x^{2} + \left[\dot\alpha(t)-\frac{\dot\lambda(t)}{\lambda(t)}\alpha(t)\right]x,
\label{eq:phase_acc_motion2}
\end{equation}
whereas $\Phi'(y,z,t)$ accounts for the dependence on the spatial coordinates $y,\, z$, and includes also terms that depend only on time.

If one is interested in the behavior of the phase just as a function of $x$, namely for $y=z=0$ as in Sect. II, then the contribution of $\Phi'$ can be safely neglected (it amounts to a global phase).
Instead, if one is interested in the dependence of the fringe wave vector upon the density of the condensate, in principle $\Phi'$ cannot be ignored (see below).

The above Eq.~(\ref{eq:phase_acc_motion2}) shows that the phase $\Phi(x,t)$ of an expanding condensate contains two terms: a quadratic term (in the coordinate $x$), whose coefficient ${\dot\lambda(t)}/{\lambda(t)}$ depends on the trap frequency \cite{dalfovo1999}, and a linear term that depends on the center-of-mass position $\alpha(t)$ and its velocity $\dot{\alpha(t)}$.

\section{Two interfering wave packets}
\label{appendix:B}

The interference of two wave packets produces a density pattern proportional to $\cos[\Delta\Phi(x,t)]=
\cos[\Phi_{1}(x,t) - \Phi_{2}(x,t)]$. According to Eq. (\ref{eq:phase_acc_motion}), we have (modulo a global phase)
\begin{equation}
\Delta\Phi(x,t) =
\left[\delta \dot\alpha(t)-\frac{\dot\lambda(t)}{\lambda(t)}\delta\alpha(t)\right]x\equiv K_{f} x
\label{eq:deltaphi}
\end{equation}
where $\delta\alpha\equiv \alpha_{1}-\alpha_{2}$, $K_{f}$ denotes the wave vector of the interference fringes. Here we are considering a system that is invariant under parity, where the two condensates experience opposite forces; thus we have $\alpha_{1}=-\alpha_{2}\equiv\alpha$, and
\begin{equation}
K_{f} = 2\left[\dot\alpha(t)- \alpha(t)\dot\lambda(t)/\lambda(t)\right].
\label{eq:kf}
\end{equation}

When the $\Phi'$ contribution is taken into account, an extra term $\Delta \Phi' =\Phi'_1 -\Phi'_2$ arises: however, since $\Phi'_1 = \Phi'_2$, $\Delta \Phi'$ is identically zero and the present model does not capture the curvature of the interference fringes due to the density dependence of the wave vector.

\bibliography{Biblio}

\end{document}